\documentclass[letterpaper]{article} 
\usepackage{aaai24}  
\usepackage{times}  
\usepackage{helvet}  
\usepackage{courier}  
\usepackage[hyphens]{url}  
\usepackage{graphicx} 
\usepackage{natbib}  
\usepackage{caption} 
\usepackage{algorithm}
\usepackage{algorithmic}

\usepackage{newfloat}
\usepackage{listings}
\DeclareCaptionStyle{ruled}{labelfont=normalfont,labelsep=colon,strut=off} 
\floatstyle{ruled}
\newfloat{listing}{tb}{lst}{}
\floatname{listing}{Listing}
\usepackage{array}

\title{Variational Exploration Module VEM: A Cloud-Native Optimization and Validation Tool for Geospatial Modeling and AI Workflows}
\author{
    Julian Kuehnert\textsuperscript{\rm 1}, 
    Hiwot Tadesse\textsuperscript{\rm 1}, 
    Chris Dearden\textsuperscript{\rm 2},
    Rosie Lickorish\textsuperscript{\rm 3},
    Paolo Fraccaro\textsuperscript{\rm 3},
    Anne Jones\textsuperscript{\rm 3},
    Blair Edwards\textsuperscript{\rm 3},
    Sekou L. Remy\textsuperscript{\rm 1},
    Peter Melling\textsuperscript{\rm 4},
    Tim Culmer\textsuperscript{\rm 4}
}
\affiliations{
    \textsuperscript{\rm 1}IBM Research, Nairobi, Kenya\\
    \textsuperscript{\rm 2}STFC Hartree Centre, Warrington, UK\\
    \textsuperscript{\rm 3}IBM Research, Daresbury, UK\\
    \textsuperscript{\rm 4} Riskaware Ltd., Bristol, UK\\
    julian.kuehnert@ibm.com, hiwot.belay.tadesse@ibm.com, chris.dearden@stfc.ac.uk, rosie.lickorish@uk.ibm.com, paolo.fraccaro@ibm.com, anne.jones@ibm.com,  bedwards@uk.ibm.com, sekou@ke.ibm.com,
    peter.melling@riskaware.co.uk,
    tim.culmer@riskaware.co.uk
}

\usepackage{bibentry}

\begin{document}

\maketitle

\begin{abstract}
Geospatial observations combined with computational models have become key to understanding the physical systems of our environment and enable the design of best practices to reduce societal harm. Cloud-based deployments help to scale up these modeling and AI workflows. Yet, for practitioners to make robust conclusions, model tuning and testing is crucial, a resource intensive process which involves the variation of model input variables. We have developed the Variational Exploration Module which facilitates the optimization and validation of modeling workflows deployed in the cloud by orchestrating workflow executions and using Bayesian and machine learning-based methods to analyze model behavior. User configurations allow the combination of diverse sampling strategies in multi-agent environments. The flexibility and robustness of the model-agnostic module is demonstrated using real-world applications.
\end{abstract}

\section{Introduction}
Climate and environmental modeling has become an integral part of policy and decision making, not only for climate change adaptation strategies but also for responding to environmental hazards in general. Examples include (near) real-time monitoring of weather observations feeding into early-warning systems, climate impact modeling of disaster scenarios to design preventive infrastructure, and modeling of urban odour plume dispersion to inform evacuation strategies. 

Computational pipelines can consist of both physics- or AI-based modeling components. Traditional computational infrastructures limit the flexibility of these computational pipelines, which scale as the complexity of these computational workflows increases and the volume of geospatial-temporal observations grows. To overcome these data and computational challenges, and to facilitate easy access and interaction with modeling workflows for end users, a number of end-to-end cloud-based modeling platforms have been developed recently \cite{bresch2021climada, edwards2022climate}. 

For practitioners to design robust and trustworthy solutions, it is indispensable for them to understand the model behavior with regards to how input variables influence model outcomes \cite{wagener2022evaluation}. Here we present the Variational Exploration Module, referred to as VEM, which allows validation and optimization of cloud-deployed geospatial modeling workflows. This includes calibration and tuning of model (hyper-)parameters, model input sensitivity analysis, as well as uncertainty quantification. The module's framework is model agnostic and allows model workflows to be evaluated with custom metric functions in a modular way.

\section{Deployment and Architecture}
\subsection{IBM's Geospatial Platform}
IBM's\footnote{IBM and the IBM logo are trademarks of International Business Machines Corporation, registered in many jurisdictions worldwide. Other product and service names might be trademarks of IBM or other companies. A current list of IBM trademarks is available on \url{ibm.com/trademark}.} Geospatial Discovery Network, GeoDN, is a cloud-native, scalable, modular platform for geospatial modeling, data and AI.  GeoDN Modeling, originally named Climate Impact Modeling Framework (CIMF) \cite{edwards2022climate}, allows the construction of flexible, modular workflows, consisting of containers that contain simulation or AI models, data querying or processing steps, or other operations, built using any programming languages and dependencies.  It provides ease of use with onboarded models/workflows which can be triggered by a simple json specifying user options, with workflow graphs generated dynamically based on user choices.  In addition, the modular nature of the workflows means they can be easily  augmented with additional steps/functionality or organised in different configurations for different purposes (i.e. single event, ensemble, iterative calibration).  A catalogue of workflows can be utilised by users, triggering workflows through the REST APIs or Python SDK.  In the backend, the framework uses Kubeflow Pipelines+Tekton for workflow execution and S3-compatible object storage as the data exchange
layer; it can be deployed on any OpenShift cluster.

\subsection{System Architecture}
\begin{figure*}[t]
\centering
\includegraphics[width=0.95\textwidth]{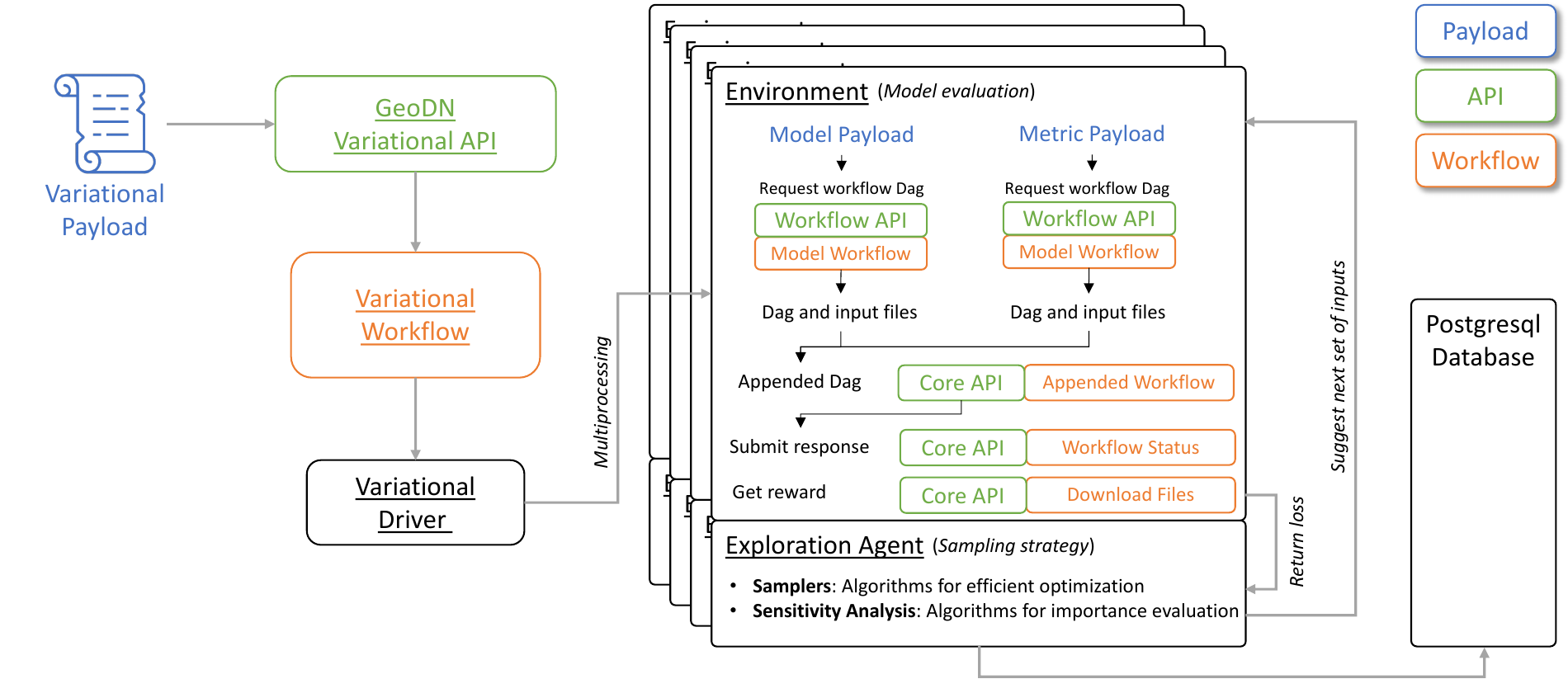} 
\caption{System architecture of the Variational Exploration Module VEM deployed on the OpenShift container registry. As indicated on the top right, payloads, APIs and workflows are color-coded in blue, green, and orange, respectively.}
\label{fig:system}
\end{figure*}

\begin{figure}[t]
\centering
\includegraphics[width=0.95\columnwidth]{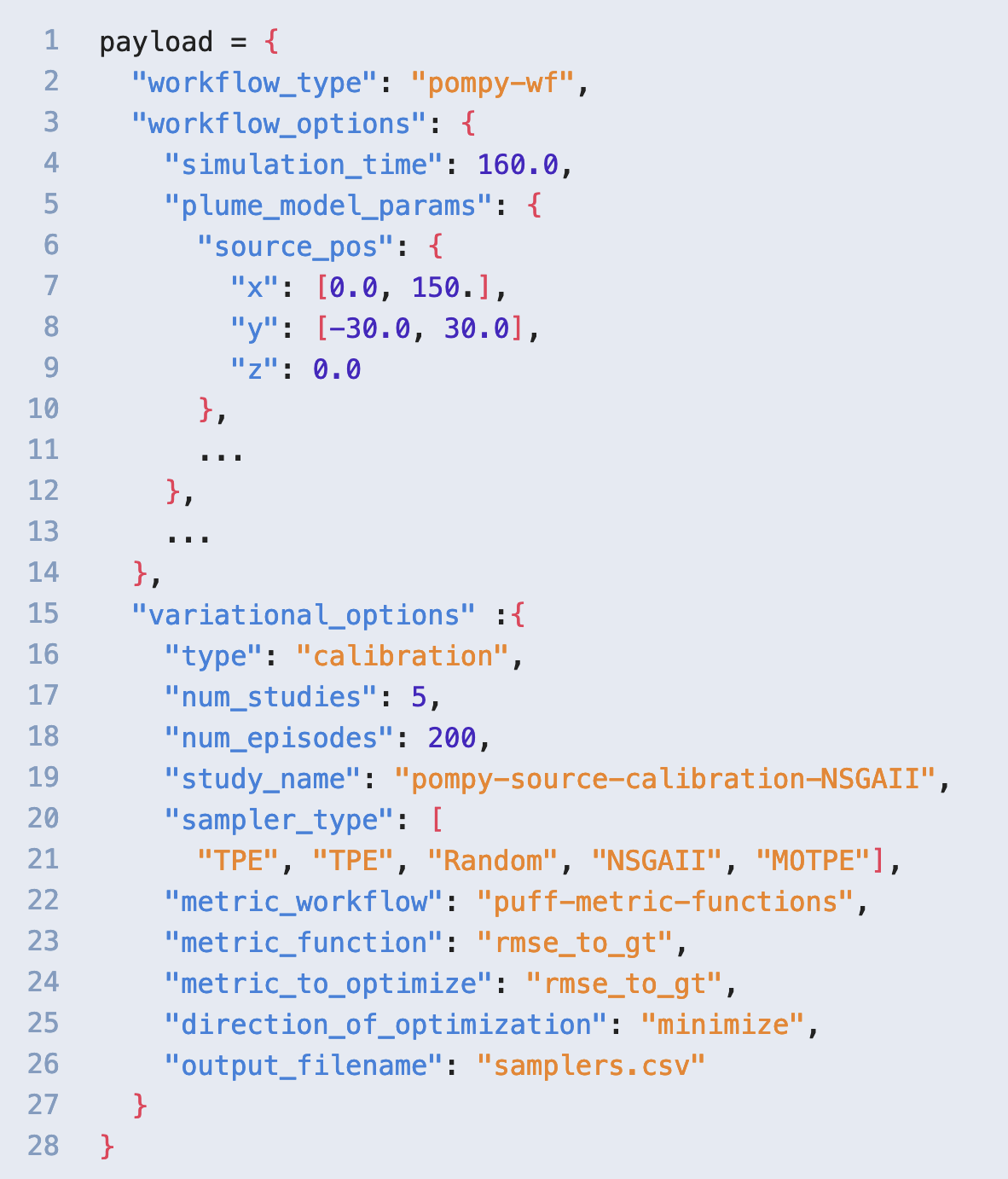} 
\caption{Example payload for the Pompy workflow as presented in the experimental results. The workflow to be studied is specified in \texttt{workflow\_type}, while workflow parameters are specified in \texttt{workflow\_options}, with variable ranges in square brackets. Study configurations are given in \texttt{variational\_options}, specifying the type of study such as calibration or sensitivity analysis. \texttt{num\_studies} specifies the number of studies in parallel with \texttt{num\_episodes} of iterations each. All parallel studies exchange information via a common database of unique name unless a list of different names is given in \texttt{study\_name}. Different sampler types can be assigned for the parallel studies. For multi-objective optimization, \texttt{metric\_to\_optimize} takes a list of evaluation metrics, for which the directions of optimization can also be specified in a list.}
\label{fig:payload}
\end{figure}

VEM is provided as a microservice behind GeoDN's Variational API. The full system architecture is illustrated in Figure \ref{fig:system}. 
Payloads submitted to the Variational API define the name of the workflow to be studied (\verb|workflow_type|), the workflow parameters including value ranges thereof (\verb|workflow_options|), as well as options which define the type of study to be carried out (\verb|variational_options|). An example payload is shown in Figure \ref{fig:payload}. 
The submission of a payload to the Variational API opens a workload pod on the container platform in which the the Variational Driver is launched. The Driver orchestrates the user-defined study by initiating pairs of Environments and Exploration Agents. Multiple pairs can be launched in parallel within a multiprocessing pool\footnote{\url{https://docs.python.org/3/library/multiprocessing.html}}. 
The Exploration Agent initiates an Optuna study object \cite{akiba2019optuna} which manages all model evaluation samples as so-called trials. The Agent is assigned a sampler as detailed in the payload (\verb|sampler_type|) that defines the strategy of how the parameter space is sampled. This way, the Agent iteratively suggests new parameter sets to be evaluated by the Environment. After model evaluation, the Environment returns the reward/loss value to the Agent, which stores all input parameters and corresponding loss values in a global PostgreSQL\footnote{\url{https://www.postgresql.org/}} database.

\subsection{Workflow Orchestration}
For a model to be evaluated with regards to a specific input parameter set, the Environment pulls the workflow definitions for both the model and the metric function from GeoDN's Workflow API in the form of directed-acyclic graphs (DAGs). The two graphs are appended to each other and submitted as a combined workflow. This modular approach gives users the flexibility to combine deployed models with custom metric evaluation functions by simply specifying the appropriate workflow names in the payload. After completion of the pipeline run, which is verified by monitoring the workflow status via GeoDN's Core API, the reward/loss value is downloaded and passed to the Exploration Agent.

\subsection{Input Parameter Space}
To define the parameter space to be examined, the user includes ranges in the payload under the workflow options that contain all workflow inputs. Ranges can be defined in the following forms:   
\begin{itemize}
    \item continuous: \texttt{[float1, float2]},
    \item discrete: \texttt{[int1, int2]},
    \item logarithmic: \texttt{[float1, float2, "log"]},
    \item categorical: \texttt{["cat1", "cat2", ...]}.
\end{itemize}
The continuous and logarithmic ranges are defined by floating-point numbers, where \texttt{float1} $<$ \texttt{float2}, the discrete range is defined by integers, where \texttt{int1} $<$ \texttt{int2}, and the categorical range is defined by strings. 

\section{Framework Capabilities}
\subsection{Hybrid Cloud Multi-Agent Optimization}
As mentioned earlier, the framework allows the creation of multiple optimization agents to run studies in parallel. The user can define the number of studies in the payload (\verb|num_studies|). If all studies are given the same name (\verb|study_name|), their trials are written to the same database. In this way, the agents exchange all intermediate results with each other and can exploit the accumulated prior knowledge to suggest new parameter sets. The same principle of defining the same study name can be leveraged to resume interrupted studies or to run optimization agents from different server or cloud instances, assuming that the PostgreSQL database can be accessed from each instance. 

\subsection{Sampling Strategy Diversification}
The efficiency of sampling strategies is inevitably problem dependent. This assertion, formalized in No-Free-Lunch theorems \cite[e.g.][]{wolpert1997no, fichtner2019hamiltonian}, states that there is no universally ideal optimization method. To overcome this and increase the robustness of black-box optimization, VEM allows ensembling and learning across different sampler types. At present, samplers provided by the hyperparameter optimization framework Optuna\footnote{\url{https://optuna.readthedocs.io/en/stable/reference/samplers/}} are supported.

\subsection{Computational Performance Metrics}
VEM allows users to evaluate workflows based on computational performance metrics. The process by which these metrics are extracted from OpenShift pods is as follows. Once the model processing has completed for a given experiment, the performance metrics are extracted by running an additional step within the pod. A HTTP GET request is sent to the cluster's internal monitoring server, where a comprehensive collection of metrics data from across the entire environment is collated and stored. The query requests information on CPU, memory usage and i/o bandwidth for the past 24 hours, at 10 second intervals, to sufficiently capture the performance of the experiment during this time. Once collected, the metrics are saved to a CSV file, along with a set of aggregated statistics, and uploaded to cloud object storage. VEM pulls these metrics and adds them to the Optuna study object under the corresponding trial.

On Openshift, we take advantage of Prometheus as the cluster's internal monitoring server, an open-source systems monitoring and alerting toolkit that collects and stores cluster wide metrics \cite{208870}. Prometheus can be used with other cloud providers, as could alternative monitoring software. Prometheus is widely used and a common choice across varies cloud environments.

\subsection{Multi-Objective Optimization and Sensitivity Analysis}
To perform multi-objective optimization, the user can specify a list of metric functions to be optimized, which can be any combination of model evaluation and computational performance metrics. If the list contains multiple entries, a list of the same length is given to the directions of optimization.  \\
To evaluate the influence of each input variable on the workflow output, sensitivity analysis can be performed by specifying the type of study in the variational options of the payload. This triggers a calculation of the relative parameter importances after each completion of a series of 20 trials (globally across multiple exploration agents, if applicable). The importances are calculated with respect to the metric function value, using by default the Functional Analysis of Variance (fANOVA) evaluation algorithm \cite{Hutteretal2014} as implemented by Optuna. The algorithm trains a random forest regression model that predicts the objective values of completed trials given their input parameter configurations. The returned sensitivity indices describe the relative parameter importances on a scale between zero and one and are of first order, that is, the fractional variance resulting from each parameter averaged over all instantiations of the other parameters. The sensitivity values are stored in the form of ordered dictionaries under user attributes of the Optuna study object and their evolution as a function of number of iterations can be monitored to check convergence.

\section{Experimental Results}
In the following, we show experimental results focusing on parameter calibration and sensitivity analysis of atmospheric dispersion models motivated by a collaboration between IBM, STFC, and the UK-based incident modelling consultancy company Riskaware. Previously, VEM was used to perform uncertainty quantification with a pluvial flood model by \citet{jones2023ai} in which uncertainties in climate drivers and input parameters are jointly projected to maximize the spread of possible flood scenarios. The collection of results, in addition to hyperparameter tuning, for which the underlying Optuna samplers are a popular choice, demonstrate the versatile capabilities of VEM for real-world applications.

\subsection{Source Calibration in Odour Plume Model}
In this section, the calibration capability of VEM is used to find the emission source location of an odour plume. The deployed model is PomPy \cite[Puff-based odour plume model in Python,][]{farrell2002filament}, an open-source atmospheric dispersion model which allows 2D odour concentration fields driven by an underlying dynamic wind field to be calculated.

The model is used to generate plume concentration time series measured at 20 randomly located sensor locations from a hypothetical single point emission source and a noisy wind field. Subsequently, the hypothetical source position is found by sampling the source location coordinates within the entire domain of simulation, under the assumption that the wind field is known. The goodness of the proposed source location is measured by the misfit of the measured time series. The metric used to compare the simulated concentration time series to the hypothetical ground truth time series is the average of the root mean square error (RMSE) across all sensor locations.

\begin{figure}[t]
\centering
\includegraphics[width=0.95\columnwidth]{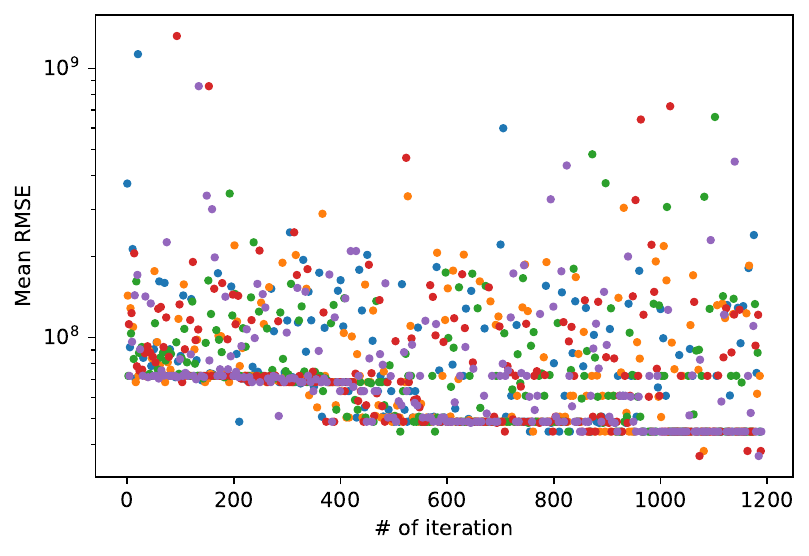} 
\caption{
Convergence of the sampling algorithm with increasing number of iterations by minimizing the misfit function, here the mean RMSE, to find the best source location. The dots in five different colors represent samples from the five parallel studies which were sharing information. Progressive global minimization demonstrates the efficiency of the information exchange for optimization.}
\label{fig:pompy_convergence}
\end{figure}

In this experiment, the NSGAII sampler is used in five parallel studies to strategically minimize the misfit between ground truth and predicted concentration time series and hence to find the source coordinates. Figure \ref{fig:pompy_convergence} shows the metric value, i.e., the mean RMSE across all sensor locations, as a function of the number of iterations. 
It can be seen how the NSGAII samplers balance exploration and exploitation to search the parameter space and minimize the misfit function progressively and globally across the five parallel studies that share all samples which each other to increase prior knowledge for more efficient optimization.

\begin{figure*}[t]
\centering
\includegraphics[width=0.95\textwidth]{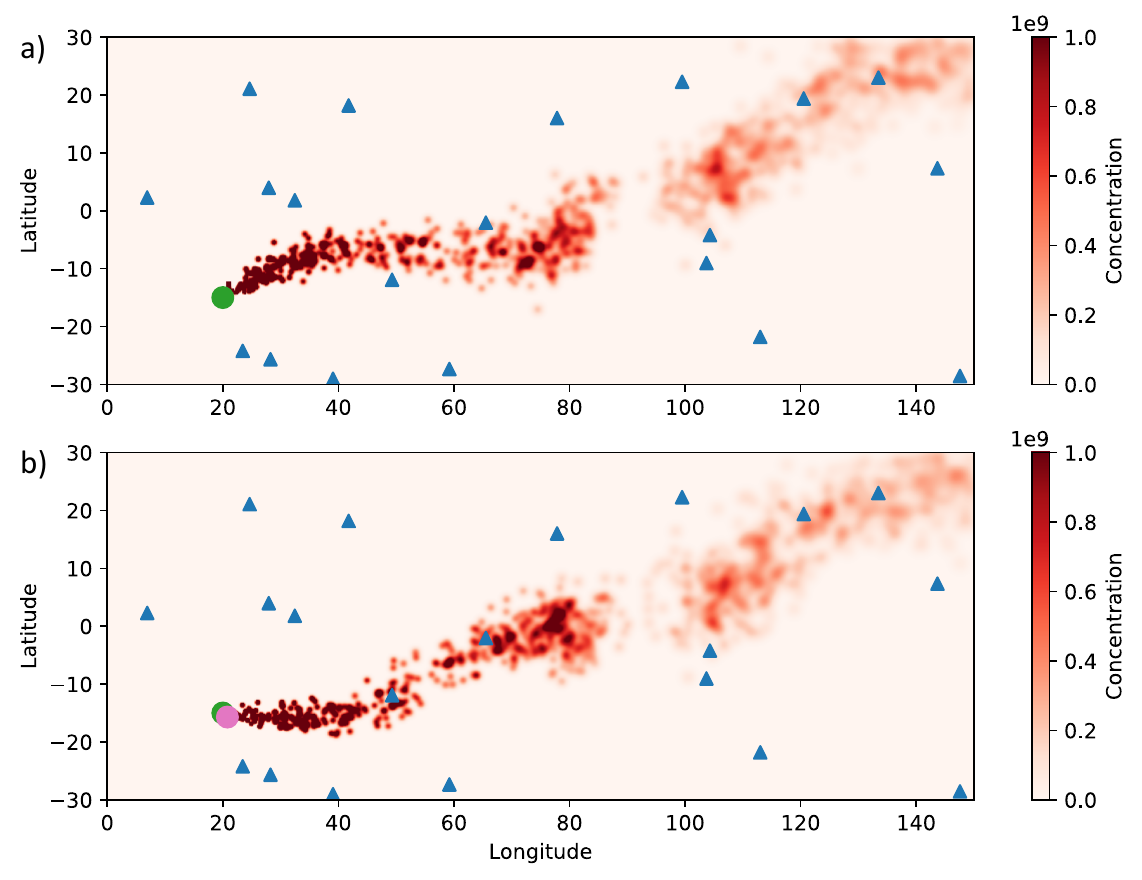} 
\caption{A snap shot of the odour concentration fields corresponding to a) the source location of the ground truth (green dot) and b) the source location identified as the best candidate by the calibration algorithm (pink dot). While the overall dispersion paths of the plumes are similar, differences in the contours are visible. Blue triangles mark the positions of the 20 sensors that measure concentration time series. }
\vspace{-0.5cm}
\label{fig:pompy_plume}
\end{figure*}

Figure \ref{fig:pompy_plume} shows the odour concentration field as generated by a) the source location of the ground truth compared to b) the source location identified as the best candidate by the calibration algorithm. 
The comparison of the two concentration fields shows that the overall dispersion paths of the plumes are similar, yet differences in the contours are visible. This means that despite the well reconstructed source location, the small residual difference has a large impact on the resulting concentration field. This can be mainly attributed to the noise in the wind field, which makes the calibration very difficult. This could explain why up to 1000 iterations were required despite the comparatively small parameter space of two dimensions.

\subsection{Sensitivity Analysis in Urban Dispersion Model}

In this section we present results from a sensitivity analysis performed on the Urban Dispersion Model (UDM) and automated using VEM. UDM is a Gaussian puff model developed by Riskaware, a UK-based incident modelling consultancy company, and is used to estimate the downwind dispersion of airborne contaminants in rural and urban areas. The sensitivity analysis was designed to identify which UDM parameters have the most significant impact on computational memory usage, and to isolate any outlier cases where the memory usage is unusually high. 

For the experimental design, VEM performed a total of 976 experimental trials of UDM for a geographic location covering the city of Newcastle in North-East England, UK. Each UDM trial was configured to run for a fixed simulation time of 20 minutes, with the source term emitting continuously at 1 second intervals for the duration of the simulation period. For each trial, the release location was chosen at random within a bounding box encompassing the city limits. Including the latitude and longitude coordinate of the release location, a total of 13 model input parameters were selected for the study, each sampled randomly from a specified range of values, and covering a mixture of meteorological factors and physical factors that define the properties of the simulated atmospheric plumes. The metric for which the sensitivity indices are calculated is the maximum memory usage within the Kubernetes pod running the model. 

\begin{figure}[t]
\centering
\includegraphics[width=0.95\columnwidth]{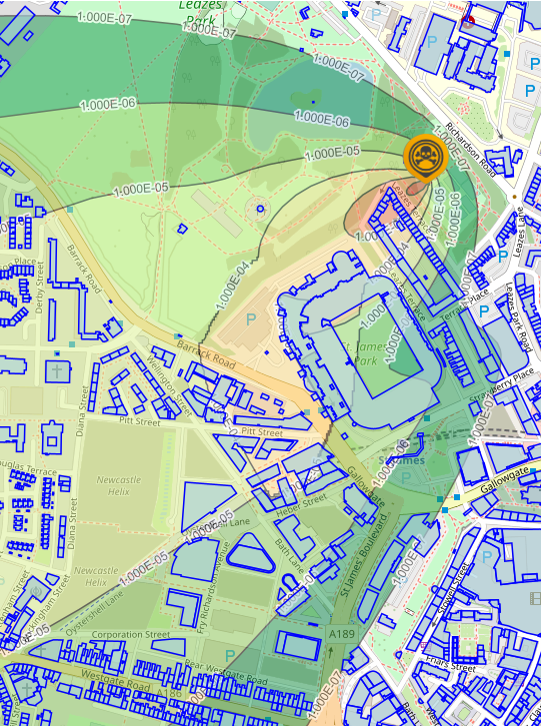} 
\caption{Map showing the release location (orange symbol) and resultant spread of gaseous material (coloured contours) in the context of the surrounding urban topology for an UDM experimental with particular high memory usage. VEM enables to easily include computational performance metrics into the optimization algorithm for practitioners to jointly interpret computational and physical properties. Contours represent dosage of released material in $kg s m^{-3}$.}
\label{fig:udm_outlier_case}
\end{figure}

Table~\ref{tab:udm_sensitivities} shows the parameter importances calculated using the fANOVA-based algorithm, and reveals that the latitudinal coordinate of the release location has the largest impact on memory usage, followed by the longitudinal coordinate. 
\begin{table}
\begin{tabular}{| c | c | c |}
\hline
  \textbf{Input category} & \textbf{Parameter name} & \textbf{Importance} \\
\hline
  source term & latitude & 0.4385 \\
  source term & longitude & 0.158 \\  
  meteorology & windspeed & 0.0844\\
  material & density & 0.0655  \\
  meteorology & wind direction & 0.0566  \\
  meteorology & temperature & 0.0557  \\
  material & deposition velocity & 0.0479  \\
  source term & mass & 0.0336  \\
  puffSplitting & shearFactor2 & 0.0286  \\
  puffSplitting & windDivSplitAngle & 0.0122  \\
  puffSplitting & groundOffsetSplitSize & 0.0082  \\
  puffSplitting & groundSplitAspectRatio & 0.0073  \\
  puffSplitting & windSplitAspectRatio & 0.0036  \\
\hline
\end{tabular}
\caption{UDM parameter importances with respect to peak memory usage, calculated using the fANOVA method of~\citet{Hutteretal2014}.}
\label{tab:udm_sensitivities}
\end{table}
It is also apparent from table~\ref{tab:udm_sensitivities} that the puffSplitting parameters have the least impact on memory usage. From the PostgreSQL database containing the results from each experimental trial, the distribution of memory usage across all trials were analysed to identify outlier cases. We could then perform further analysis on these outlier cases to investigate the possible causes of the high memory usage. Figure~\ref{fig:udm_outlier_case} shows a map view from an experimental trial where the maximum memory usage peaked at 3.7GB (3 standard deviations from the ensemble mean). 
The contours in the plot show the simulated dosage associated with the spreading of the atmospheric plume, with the footprint of the local buildings outlined in blue. The figure suggests that some complex interactions between the puffs and the nearby urban structures are taking place which could account for the relatively high memory usage. For instance, the incoming wind direction for this trial is north-easterly (58.59 degrees) and so puffs are advected towards the row of terraced housing which are within 10-20$m$ of the release location and arranged almost perpendicular to the incoming flow. Immediately behind the row of terraces is a football stadium, which presents another obstacle for the puffs to interact with. The wind speed is also above average for this case (5.45$ms^{-1}$) which could be amplifying the turbulent interactions with buildings in the immediate vicinity. 

This real-world application shows that VEM facilitates the inclusion of computational performance metrics in the optimization process. In this case, this was very insightful as it allowed outliers to be identified and the physical reasons for them to be explored. Since VEM also supports multi-opjective optimization, it is possible to optimize the model evaluation and computational performance metrics together.


\section{Conclusion and Future Work}
In this paper, we present the Variational Exploration Module VEM for input space optimization and sensitivity analysis of geospatial modeling and AI workflows. As implemented, the module supports workflows deployed as Kubeflow pipelines on Kubernetes container platforms and is able to perform hybrid-cloud multi-agent optimization as well as ensembling and learning across different sampler types. Motivated by a collaboration between STFC and IBM under the Hartree National Centre for Digital Innovation, and the UK-based incident modelling consultancy company Riskaware, experimental results for parameter calibration and sensitivity analysis of atmospheric dispersion models are shown. 

In future, we plan to expand the number of sampling algorithms available for strategic exploration of the input space and the number of methods available for sensitivity analysis. This includes implementing samplers based on reinforcement learning to support state variables. We are also working on an improved API and user interface to provide users with informative statistics and graphs that summarize the experimental results. 

\section{Acknowledgments}
This work was supported by the Hartree National Centre for Digital Innovation, a collaboration between STFC and IBM.

\bibliography{aaai24}

\end{document}